\documentclass[fleqn,10pt]{SelfArx} 

\usepackage[english]{babel} 

\usepackage{lipsum} 
\usepackage{afterpage}

 \definecolor{color1}{RGB}{0,0,110} 
\definecolor{color2}{RGB}{0,20,80} 

\usepackage{hyperref} 
\hypersetup{hidelinks,colorlinks,breaklinks=true,urlcolor=color2,citecolor=color1,linkcolor=color1,bookmarksopen=false,pdftitle={Title},pdfauthor={Author}}

\JournalInfo{Uploaded to arXiv.org} 
\Archive{} 

\PaperTitle{Multiscale analysis of the effect of surface charge pattern on a nanopore's rectification and selectivity properties: from all-atom model to Poisson-Nernst-Planck} 

\Authors{
M\'onika Valisk\'o\textsuperscript{1}*, 
Bart\l{}omiej Matejczyk\textsuperscript{2} ,  
Zolt\'an Hat\'o\textsuperscript{1},
Tam\'as Krist\'of\textsuperscript{1},
Eszter M\'adai\textsuperscript{1,2},  
D\'avid Fertig\textsuperscript{1},
Dirk Gillespie\textsuperscript{4},
Dezs\H{o} Boda\textsuperscript{1}
} 
\affiliation{\textsuperscript{1}\textit{Department of Physical Chemistry, University of Pannonia,  P. O. Box 158, H-8201 Veszpr\'em, Hungary}} 
\affiliation{\textsuperscript{2}\textit{Department of Mathematics, University of Warwick, CV4 7AL Coventry, United Kingdom}}
\affiliation{\textsuperscript{3}\textit{Department of Material- and Geo-Sciences, Technische Universit\"{a}t  Darmstadt, Petersenstr.\ 23, D-64287 Darmstadt, Germany}}
\affiliation{\textsuperscript{4}\textit{Department of Physiology and Biophysics, Rush University Medical Center, Chicago, IL 60612, USA}}
\affiliation{*\textbf{Corresponding author}: valisko@almos.vein.hu} 

\newcommand{\keywordname}{Keywords} 

\Abstract{We report a multiscale modeling study for charged cylindrical nanopores using three modeling levels that include (1) an all-atom explicit-water model studied with molecular dynamics (MD), and reduced models with implicit water containing (2) hard-sphere ions studied with the Local Equilibrium Monte Carlo simulation method (computing ionic correlations accurately), and (3) point ions studied with Poisson-Nernst-Planck (PNP) theory (mean-field approximation). 
We show that reduced models are able to reproduce device functions (rectification and selectivity) for a wide variety of charge patterns; that is, reduced models are useful in understanding the mesoscale physics of the device (i.e., how the current is produced). 
We also analyze the relationship of the reduced implicit-water models with the explicit-water model and show that diffusion coefficients in the reduced models can be used as adjustable parameters with which the results of the explicit- and implicit-water models can be related. 
We find that the values of the diffusion coefficients are sensitive to the net charge of the pore, but are relatively transferable to different voltages and charge patterns with the same total charge.
}

\begin{document}

\flushbottom 
\maketitle 

\section{Introduction}
\label{sec:intro}

Nanopores make the transport of ions (or, more generally, the fluid) through membranes possible \cite{albrecht_chapter_2013,iqbal_book_2011}.
They have the special property, which distinguishes them from micropores, that their radius is measurable to the characteristic screening length of the electrolyte \cite{kuo_l_2001,vandenberg_csr_2010,singh_jpcb_2011}.
In this case, depletion zones of ions can form at the charged surfaces making the control of ion flow possible \cite{karnik_apl_2006,karnik_nl_2005,Kalman_BJ_2009,kalman_abac_2009,Kim_2015}. 

Modification of the surfaces with functional molecules facilitates designing nanodevices which can respond to single or dual stimuli \cite{zhang_cc_2013}. 
Stimuli can be voltage, pH, molecules, ions, light, or temperature or combination of them.  
This makes nanopores promising candidates to be the core units of sensors \cite{sexton_jacs_2007,gyurcsanyi_trac_2008,howorka_csr_2009,vlassiouk_jacs_2009,piruska_csr_2010,shi_ac_2016,ensinger_2018}.
Natural nanopores (called ion channels) play a crucial role in life due to their ability to facilitate controlled movement of ions through membranes separating cell compartments \cite{Hille,miedema_nl_2007}.

The molecular dimensions of both natural and synthetic nanopores require their modeling on the molecular level provided that we want to understand the molecular mechanisms behind device function and to use that knowledge to design devices with novel properties \cite{park_mfnf_2015}.
Device approach is evidently relevant in biology too \cite{eisenberg_jml_2018}.

The rise of nanopores cannot be separated from their application: their theoretical understanding and technological usage developed hand in hand \cite{Abgrall_2008,vandenberg_csr_2010,guan_nanotech_2014,zhang_nanotoday_2016}. 
Therefore, their modeling primarily focused on computing device functions (ionic currents as responses to the aforementioned stimuli) with the intention of a broad understanding the device mechanisms.
Continuum descriptions such as the Poisson-Nernst-Planck (PNP) and the Navier-Stokes theories were routinely and successfully applied to achieve this goal \cite{cervera_jcp_2006,constantin_pre_2007,ramirez_jcp_2007,kalman_am_2008,gracheva_acsnano_2008,singh_jap_2011,singh_jpcb_2011,cervera_ea_2011,apel_nt_2011,burger_nonlin_2012,Ali_ACSnano_2012,tajparast_bba_2015,perezmitta_pcp_2016,Ali_Lang_2017,Ali_AC_2018}.

As dimensions of nanopores shrink, however, the demand for models with more details rose.
This means all atom models, where not only the ions, but membrane atoms and water molecules are modeled explicitly and simulated with the molecular dynamics (MD) technique \cite{cruzchu_jpcc_2009,aksimentiev_ieee_2009,cruzchu_2009,gamble_jpcc_2014,ge_ms_2016,luan_nl_2019}.
These models provide insight into important local microscopic mechanisms, this modeling level, however, is still too expensive to simulate global device-level phenomena.

There seems to be a gap between all-atom MD simulations and device-level continuum calculations.
Filling this gap requires a multiscaling approach \cite{steinhauser_multiscale_2008} that includes models of varying resolutions and computational methods with which we study these models.
Models with less resolution are called reduced models.
The expression ``coarse-graining'' is also widespread, but we prefer the word ``reduced'' because it reflects the fact that the degrees of freedom that are explicitly modeled are reduced by averaging them into various response functions.
The most trivial example for this is the implicit-solvent model, where the solvent molecules (generally, water) are replaced by a dielectric continuum that also exerts friction on diffusing ions.

Most importantly, this multiscaling approach includes relating these modeling levels to each other.
Establishing these relations means understanding the effect of ignored degrees of freedom and finding computational links between the models.
These links are often parameters of the response functions: the dielectric and diffusion coefficients, for example.

In our previous papers \cite{hato_pccp_2017,matejczyk_jcp_2017}, we used a bipolar nanopore as a test case for this type of multiscale study.
The common point between the two studies were a hybrid method in which the Nernst-Planck (NP) transport equation is coupled to a particle simulation method, the Local Equilibrium Monte Carlo (LEMC) technique (NP+LEMC) \cite{boda_jctc_2012,boda_jml_2014,boda_arcc_2014}.
LEMC is the adaptation of the Grand Canonical Monte Carlo (GCMC) method to a non-equilibrium situation using a space-dependent chemical potential. 
LEMC is able to consider ions with finite sizes in an implicit solvent and to compute all their correlations that are beyond the mean-field level of PNP.

\begin{figure*}[t!]
\begin{center}
\includegraphics*[width=0.8\textwidth]{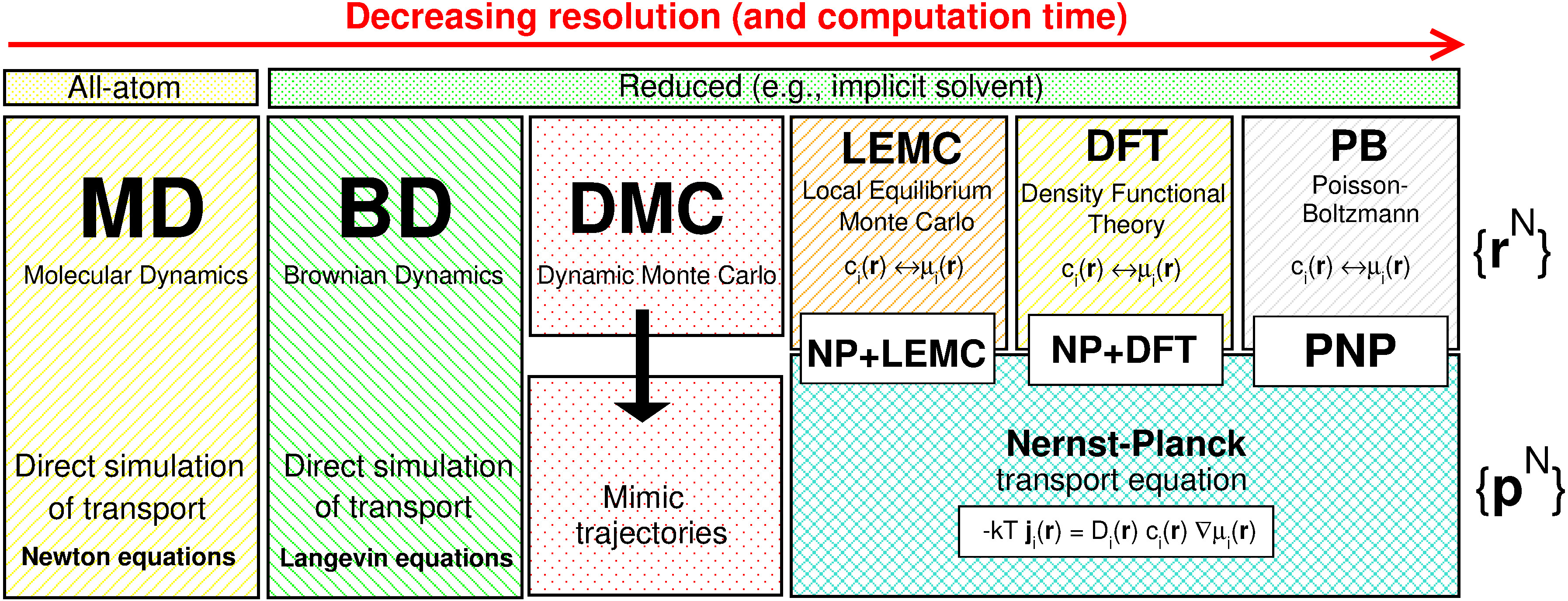}
\end{center}
\caption{Hierarchy of models and associated methods for simulating ion transport through pores. Resolution of models decreases from left to right, while methods used for studying them become more approximate, but faster. Transport can be simulated directly via treating the $\{\mathbf{p}^{N}\}$ degrees of freedom explicitly via solving equations of motion, either Newton (MD) or Langevin (BD). MC methods work over the configurational space ($\{\mathbf{r}^{N}\}$), providing either mimicked trajectories (DMC) or a relation  between chemical potential profile, $\mu_{i}(\mathbf{r})$, and concentration profile, $c_{i}(\mathbf{r})$. That relation can also be computed with theories that can be either mean-field (PB) or more state-of-the-art (DFT). If the $c_{i}(\mathbf{r})$ vs.\ $\mu_{i}(\mathbf{r})$ relation is available, it can be used in the NP equation yielding a fast technique at the price of lost details about dynamics.
}
\label{fig1}
\end{figure*}

In Ref.\ \cite{hato_pccp_2017}, we studied the effect of using implicit solvent instead of explicit solvent by comparing results of NP+LEMC to MD data.
We found that the implicit-solvent model can reproduce device function because it can reproduce the axial behavior of concentration profiles that are the main determinants of the device behavior (i.e., current) via appearance of ionic depletion zones in given axial zones of the pore as a response to applied voltage and surface charges.

In Ref.\ \cite{matejczyk_jcp_2017}, we compared NP+LEMC to PNP and analyzed the effect of using an approximate mean-field theory instead of the accurate LEMC simulations. 
We found that PNP can reproduce device behavior qualitatively because the device behavior is chiefly governed by mean-field electrostatic effects, e.g., interaction of ions with fixed pore charges and applied field.
This was confirmed by our other study for a nanopore transistor \cite{madai_pccp_2018}.

In this study, we step beyond the bipolar pore and consider various surface charge patterns.
Surface charge is an essential feature of nanopores with which selectivity and conduction properties of the pore can be influenced \cite{kuo_l_2001,Siwy_2004,yameen_nl_2009,jiang_pre_2011,zhang_acsnano_2015,singh_sab_2016}.
Surface charge can be manipulated by chemical modifications \cite{miedema06,yameen_jacscomm_2009,ali_cc_2015,ali_acsami_2015,lepoitevin_rscadv_2016}, adding components that bind selectively to functional molecules (sensing) \cite{Ali_JACS_2011,ali_cc_2015,Ali_Lang_2017,Ali_AC_2018}, or by applying voltage at embedded electrodes\cite{burgmayer_jacs_1982,karnik_nl_2005,karnik_nl_2007,kalman_am_2008,Kalman_BJ_2009,kalman_abac_2009,Fuest_nl_2015}.
Different charge patterns allow different functions of the nanopore.

In this study, we divide the nanopore's surface into two regions along the tangential axis and change the lengths and surface charge densities of those regions in a systematic way.
In this way, we can change the nanopore between two limits that are (1) symmetric and selective either for cation or anion and (2) asymmetric and rectifying.
We study both bipolar and unipolar nanopores \cite{karnik_nl_2007,vlassiouk_nl_2007,nguyen_nt_2010,tajparast_bba_2015,singh_sab_2016,huang_advfuncmat_2018}.

The MD simulations serve as the gold standards in our study, while the NP+LEMC results are computed by fitting the diffusion coefficient in the pore to MD data.
We fit these values only for a subset of all the states: bipolar nanopores in the ON state (forward biased).
Then, we examine whether these values are transferable to other states, such as unipolar nanopores, OFF states (reverse biased).
We also examine whether the diffusion coefficient values fitted for NP+LEMC are usable in PNP.
\section{Models and methods}
\label{sec:models}

\subsection{Hierarchy of models and methods}

Figure \ref{fig1} shows a series of methods that can be used to study ion transport through nanopores or ion channels.
From left to right, the models decrease in resolution with a large change from all-atom models to reduced models (with an associated decreasing trend in computation time), where solvent becomes implicit.

Transport can be simulated directly using MD, Brownian Dynamics (BD)\cite{chung-bj-77-2517-1999,im_bj_2000,berti_jctc_2014,Berti_chapter,alavizargar_jpcb_2018}, or Dynamic Monte Carlo (DMC)\cite{rutkai-jpcl-1-2179-2010,csanyi-bba-1818-592-2012,boda-jpcc-118-700-2014}, or computed with a transport equation.
Direct simulation of ionic transport in the explicit- and implicit-water frameworks are done by MD and BD, respectively, by solving the Newton and the Langevin equations of motion, respectively.
In DMC, time is absent and  trajectories are mimicked with transport properties being reproduced on average. 

If flux is computed with a transport equation (the NP equation, in this case), statistical mechanical methods are applied only over the space of configurational coordinates, $\{\mathbf{r}^{N}\}$, and their job is to relate the chemical potential profile, $\mu_{i}(\mathbf{r})$, to the concentration profile, $c_{i}(\mathbf{r})$.
The advantage of using a transport equation is that we can more easily handle cases where direct sampling of ions crossing the pore is problematic such as the case of the bipolar nanopore, through which currents are small due to depletion zones.
This approach is faster at the price of lost details about dynamics.

The applied statistical mechanical method can be a particle simulation technique (LEMC) that is more accurate but slower, a mean-field theory (Poisson-Boltzmann, PB) that is fast but approximate, or an advanced theory (e.g., Density Functional Theory, DFT).
DFT \cite{gillespie_pre_2003} can reproduce MC results quite accurately \cite{gillespie-jpm-17-6609-2005,valisko-jpcc-111-15575-2007,voukadinova_pre_2018,voukadinova_submitted_2019} at the price of being efficient only in one dimension \cite{knepley_jcp_2010}.
When coupled with NP \cite{gillespie_jpcm_2002}, DFT was successfully used to compute conduction properties of the Ryanodine Receptor calcium channel \cite{gillespie-jpcb-109-15598-2005,dirk-mike,gillespie_bj_2008,dirk-janhavi-mike,gillespie_bj_2014}.

\begin{figure*}[t!]
\begin{center}
\includegraphics*[width=0.9\textwidth]{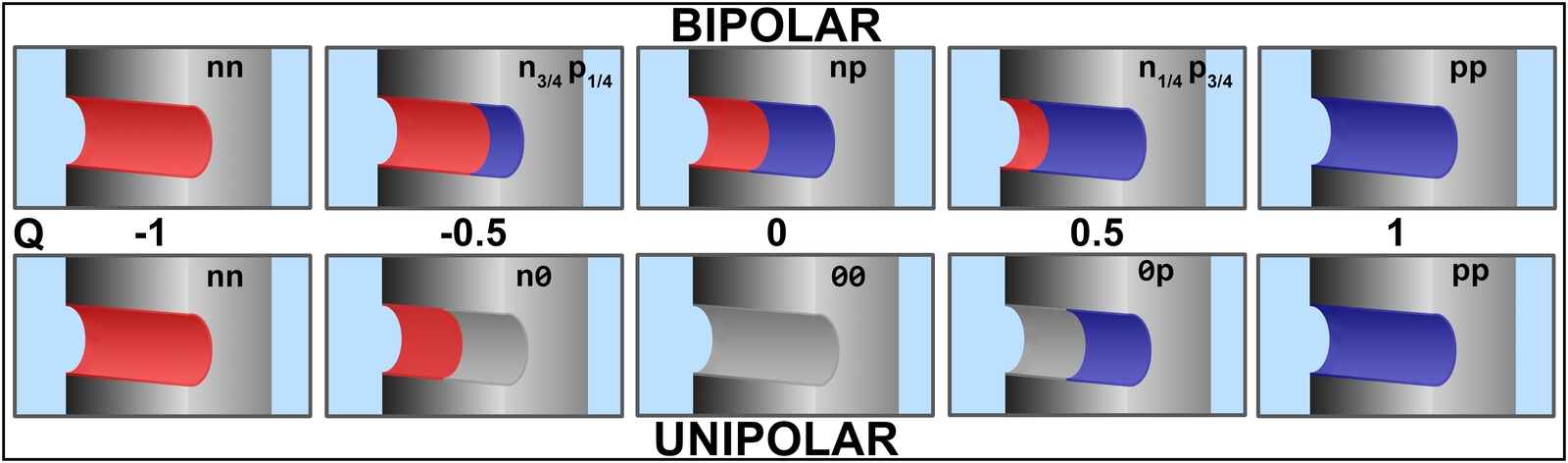}
\end{center}
\caption{Schematics of the cylindrical nanopore with different charge pattern. 
There are two regions of lengths $H_{1}$ and $H_{2}$ carrying $\sigma_{1}$ and $\sigma_{2}$ surface charges.
Both the lengths (by keeping the total length, $H=6.4$ nm fixed) and the surface charges are varied to study different kinds of pores with different  conduction properties.
We consider either bipolar (top row) or unipolar (bottom row) nanopores.
In the bipolar case, the left-hand-side region is always negative ($\sigma_{1}=\sigma_{\mathrm{n}}$, while the right-hand-side region is positive ($\sigma_{2}=\sigma_{\mathrm{p}}$) ($\sigma_{\mathrm{p}}=0.4835$ $e$/nm$^{2}$ and $\sigma_{\mathrm{n}}=-0.4835$ $e$/nm$^{2}$, in this work). 
In the unipolar case, the same is true, but the other side is neutral. 
The radius of the nanopore is $R_{\mathrm{pore}}=0.97$ nm.
The notations of the various are also indicated in the figure (for more details, see the main text).
}
\label{fig2}
\end{figure*}

In this work, we use three of these modeling levels with associated methods (MD, NP+LEMC, and PNP) and study various charge patterns for a cylindrical nanopore of fixed radius, $R_{\mathrm{pore}}$, and length, $H$.
The pores have two regions along the pore ($z$-axis) that differ in lengths ($H_{1}$ and $H_{2}$) and surface charge densities ($\sigma_{1}$ and $\sigma_{2}$).
The charge patterns are changed by changing these values (see Fig.\ \ref{fig2}).

These three methods have been presented in detail in our previous publications \cite{boda_jctc_2012,boda_jml_2014,boda_arcc_2014,hato_cmp_2016,hato_pccp_2017,matejczyk_jcp_2017,madai_jcp_2017,madai_pccp_2018}
Here, we briefly describe them focusing on the differences that are summarized in Table \ref{tab1}.

\subsection{All-atom model studied with molecular dynamics}
\label{subsec:md}

At an all-atom-resolution modeling level, the pore is formed by a carbon nanotube (CNT) between two carbon nanosheets (CNS) defining the membrane (see Fig.\ 1 of Ref.\ \cite{hato_pccp_2017}). 
Distance of CNS atoms in the two sheets (width of membrane) is $6.035$ nm, while the distance of centers of the CNT atoms from the pore axis (pore radius) is $1.136$ nm.
Placing partial charges on the C atoms of the CNT ensures that the average surface charge density corresponds to a prescribed value: either $\sigma_{\mathrm{p}}=0.4835$ $e$/nm$^{2}$, $\sigma_{\mathrm{n}}=-0.4835$ $e$/nm$^{2}$, or  $\sigma_{0}=0$  in this study. 
The model membrane was constructed with the help of the Nanotube Builder plugin of the VMD program package \cite{vmd}. 

The CHARMM27 force field \cite{charmm} was used for the $\sim$11,000 SPC water molecules (explicit water), 110-190 Lennard-Jones (LJ) and point charge type anions/cations, and LJ atoms for the CNT and CNS. 
LJ parameters and partial charges for Na$^{+}$, Cl$^{-}$, and C atoms and other details are found in the Appendix and the Supplementary information of Ref.\ \cite{hato_pccp_2017}.

The dimensions of the simulation box were 6$\times$5.2$\times$16.8 nm, with periodic boundary conditions (PBC) applied in all spatial directions. 
The systems were thermostated to 298.15 K by a modified version of the Berendsen (velocity rescaling) algorithm \cite{bussi_jcp_2007}.
To achieve $200$ mV potential difference between the two ends of the simulation box, a 0.012 V/nm external electric field was applied.
An ion was considered to cross the channel if it is initially at one side of the membrane and then ends at the opposite side of the membrane after propagating through the channel.

We applied the GROMACS (v.5.0.4) program suite \cite{Berendsen199543,Pronk01042013} using the leap-frog integrator with a 2 fs time step. 
During the MD simulation the CNS/CNT pore and membrane was kept rigid. 
Production runs were about 200-500 ns long.

\subsection{Reduced models of nanopore and electrolyte}
\label{reduced}

In our reduced models the solvent (water) is treated implicitly. 
Water is modeled as a continuum background that has two kinds of effect on ions: they screen the charges of ions and hinder the diffusion of ions via friction. 
The screening is an ``energetic'' effect that is taken into account by a dielectric constant, $\epsilon=78.5$, in the denominator of the Coulomb-potential acting between the charged hard spheres with which we model the ions:
\begin{equation}
 u_{ij}(r)
=\left\{
        \begin{array}{ll}
    \infty & \; \mbox{for} \; \;  r<R_{i}+R_{j}\\
        \dfrac{q_{i}q_{j}}{4\pi \epsilon_{0}\epsilon r} & \; \mbox{for} \; \; r \geq R_{i}+R_{j}  ,
        \end{array}
        \right. 
\label{eq:pm}
\end{equation} 
where $q_{i}$ is the charge and $R_{i}$ is the radius of ionic species $i$, $\epsilon_0$ is the permittivity of vacuum, and $r$ is the distance between the ions.
The hard sphere component is absent in the PNP calculations, as are electrostatic correlations beyond the mean-field.

Friction is a ``dynamic'' effect and it is taken into account by a diffusion coefficient, $D_{i}(\mathbf{r})$, in the Nernst-Planck (NP) transport equation for the ionic flux:
\begin{equation}
 -kT\mathbf{j}_{i}(\mathbf{r}) = D_{i}(\mathbf{r})c_{i}(\mathbf{r})\nabla \mu_{i}(\mathbf{r}),
 \label{eq:np}
\end{equation} 
where $k$ is Boltzmann's constant, $T=298.15$ K is temperature, $\mathbf{j}_{i}(\mathbf{r})$ is the particle flux density of ionic species $i$, $c_{i}(\mathbf{r})$ is the concentration, and $\mu_{i}(\mathbf{r})$ is the electrochemical potential profile.

\begin{table*}[t!]
\centering
\caption{Outline of the differences between the three models/methods.}
\begin{tabular}{|l|c|c|c|}
\hline
  & \textbf{MD} & \textbf{NP+LEMC} & \textbf{PNP} \\ \hline \hline
\textbf{Ions} &	LJ + point charges & Charged hard spheres & Point charges (mean field) \\ \hline
\textbf{Water} &	Explicit (SPC) & \multicolumn{2}{c|}{Implicit (continuum: $\epsilon$ and $D_{i}$)}\\ \hline
\textbf{Membrane} & CNS (LJ) & \multicolumn{2}{c|}{Hard wall} \\ \hline
\textbf{Pore} &	CNT (LJ + point charges) & \multicolumn{2}{c|}{Hard wall cylinder + point charges} \\ \hline \hline
\textbf{Dynamics} & Direct simulation& \multicolumn{2}{c|}{NP transport equation} \\ \hline
\textbf{Sampling} & Solving Newton's equations & Stochastic & Continuum theory \\ \hline
\textbf{Comp.\ time} & days/weeks & hours/days & seconds/minutes \\ \hline
\end{tabular} 
\label{tab1}
 \end{table*}

The diffusion coefficient profile, $D_{i}(\mathbf{r})$, is a user-specified parameter.
In the baths, we can use experimental data ($D_{\mathrm{Na}^{+}}^{\mathrm{bath}}=1.334 \times 10^{-9}$ m$^{2}$s$^{-1}$ for Na$^{+}$ and $D_{\mathrm{Cl}^{-}}^{\mathrm{bath}}=2.032 \times 10^{-9}$ m$^{2}$s$^{-1}$  for Cl$^{-}$), but inside the pore it can serve as an adjustable parameter (for the procedure of adjustment, see the Results section).
We can adjust these values to experiments as we did in the case of calcium channels \cite{boda_arcc_2014,fertig_mp_2018} or to results of MD simulations as we did previously \cite{hato_pccp_2017} and as we do in this study.

The cylindrical pore's radius and length were calculated to mimic the CNT model of the MD simulations as closely as possible on the basis of an estimated distance of closest approach of ions to the carbon atoms. 
We used the values $R_{\mathrm{pore}} = 0.97$ nm and $H = 6.4$ nm for the pore radius and length, respectively. 
The fractional point charges have been placed at the same positions as in the CNT model.

The ionic radii are the Pauling radii:  $R_{\mathrm{Na}^{+}}=0.095$ and $R_{\mathrm{Cl}^{-}}=0.181$ nm for Na$^{+}$ and Cl$^{-}$, respectively.
The size of the simulation cell is similar to that used in MD, but it is closed (no PBC applied) and cylindrical-shaped. 
Boundary conditions for the electrochemical potential are prescribed on that cylinder.
The system is rotationally symmetric, so variables in the NP equation are expressed as functions of the axial and radial coordinates, $(z,r)$.

\subsection{Nernst-Planck coupled to Local Equilibrium Monte Carlo}

To solve the NP equation together with the continuity equation ($\nabla \cdot \mathbf{j}_{i}(\mathbf{r})=0$), we need a closure between the concentration profile, $c_{i}(\mathbf{r})$, and the electrochemical potential profile, $\mu_{i}(\mathbf{r})$. 
It is provided by statistical mechanics. 
Here we use two methods to solve the problem, a particle simulation method (LEMC) and a continuum theory method (PNP). 

The LEMC technique \cite{boda_jctc_2012,boda_jml_2014,boda_arcc_2014} is practically a GCMC simulation devised for a non-equilibrium situation.
We divide the system into small volume elements, assume local equilibrium in them (they are characterized by constant $\mu_{i}$ values), and apply particle insertion and deletion steps with the same formula for acceptance probabilities as in equilibrium GCMC simulations, but using the local chemical potential of the volume element.

The input variable of the LEMC simulation is the chemical potential profile, $\mu_{i}(\mathbf{r})$, while the output variable is the concentration profile, $c_{i}(\mathbf{r})$. 
The LEMC method correctly computes volume exclusion and electrostatic correlations between ions, so it is beyond the mean-field level of the PNP theory that is routinely applied for nanopores \cite{cervera_jcp_2006,constantin_pre_2007,ramirez_jcp_2007,kalman_am_2008,gracheva_acsnano_2008,singh_jap_2011,singh_jpcb_2011,cervera_ea_2011,apel_nt_2011,burger_nonlin_2012,Ali_ACSnano_2012,tajparast_bba_2015,perezmitta_pcp_2016,Ali_Lang_2017,Ali_AC_2018}.

The NP equation is solved self-consistently with the LEMC simulations in an iteration procedure that ensures that the continuity equation is satisfied (the $\mu_{i}$ profile is varied during the iteration).
The resulting NP+LEMC technique provides a solution for the statistical mechanical problem (e.g., the $c_{i}$ vs.\ $\mu_{i}$ relation) on the basis of particle simulations, while it still gives an approximate indirect solution for the dynamical problem through the NP equation.

\subsection{Poisson-Nernst-Planck}
\label{subseq:pnp}

The other method with which we solve the system described in Subsection \ref{reduced} is a two-dimensional version of the  PNP  theory as described by Matejczyk et al.\ \cite{matejczyk_jcp_2017}.
As opposed to LEMC, which is a molecular simulation method, PNP is a mean field method that does not consider the particles as individual entities, but investigates the concentration profiles as the probability of finding a center of  a particle in a certain location. 
The concentration  depends on the interaction energy of the ion with the average (mean) electrical potential, $\Phi(\mathbf{r})$, produced by all the charges in the system, including all the ions. 

The difference between LEMC and PNP can be expressed by the electrochemical potential that, in the general case, is
\begin{equation}
 \mu_{i}(\mathbf{r}) = \mu_{i}^{0} + kT\ln c_{i}(\mathbf{r}) + q_{i}\Phi(\mathbf{r})  + \mu_{i}^{\mathrm{BMF}}(\mathbf{r}),
\label{eq:elchempotpnp}
\end{equation} 
 where $\mu_{i}^{0}$ is a reference term and $\mu_{i}^{\mathrm{BMF}}(\mathbf{r})$ is the term which is `beyond mean field' (BMF) absent in PNP, but computed in LEMC.
The BMF term accounts for all the two- and many-body correlations between ions and hard-sphere exclusion ignored by PNP (the BMF term and $q_{i}\Phi(\mathbf{r})$ together is  the excess chemical potential) .
The behavior of the BMF and other terms for the bipolar nanopore has been discussed in our previous work \cite{matejczyk_jcp_2017}.

The concentration profiles are related to the mean electrical potential through Poisson's equation.
Details about solution of the system (the Scharfetter--Gummel scheme \cite{gummel1964self}), the boundary conditions, and the mesh can be found in our earlier papers \cite{matejczyk_jcp_2017,madai_pccp_2018}.

\section{Results}
\label{sec:results}

We have performed two sets of simulations for bipolar and unipolar nanopores.
In both cases, the pore is divided into two regions along the $z$-axis of lengths $H_{1}$ and $H_{2}$ carrying $\sigma_{1}$ and $\sigma_{2}$ surface charges (Fig.\ \ref{fig2}).
We change the charge pattern by varying these values in a systematic way.

\subsection{Charge patterns}

In the case of bipolar nanopores (top row of Fig.\ \ref{fig2}), the left hand side region is negative ($\sigma_{1}=\sigma_{\mathrm{n}}$, red color), while the right hand side is positive ($\sigma_{2}=\sigma_{\mathrm{p}}$, blue color).
We gradually increase $H_{1}$, while keeping the total length, $H$, fixed, so the length of the right region decreases: $H_{2}=H-H_{1}$.
Therefore, we can characterize the geometry by the fraction of the left region: $x_{1}=H_{1}/H$.
The various charge patterns for the bipolar pores will be denoted by '$\mathrm{n}_{x_{1}}\mathrm{p}_{1-x_{1}}$'. 

The charge patterns in the top row of Fig.\ \ref{fig2}, for example, are denoted by '$\mathrm{nn}$', '$\mathrm{n}_{3/4}\mathrm{p}_{1/4}$', '$\mathrm{np}$', '$\mathrm{n}_{1/4}\mathrm{p}_{3/4}$', and '$\mathrm{pp}$'.
Subscripts $1/2$ will be dropped for simplicity, so we will refer to the usual bipolar pore '$\mathrm{n}_{1/2}\mathrm{p}_{1/2}$' as '$\mathrm{np}$'.
The limiting cases, the uniformly charged negative and positive pores, will be referred to as '$\mathrm{nn}$' and '$\mathrm{pp}$'.

These limiting cases are common in the bipolar and unipolar pores (left and right end cases of top and  bottom rows of Fig.\ \ref{fig2} are the same).
In the unipolar cases, the nanopore is composed of a charged and an uncharged region.
We change the charge pattern in a way that the negative region is always on the left ($\sigma_{1}=\sigma_{\mathrm{n}}$, $\sigma_{2}=\sigma_{0}$), while the positive region is always on the right ($\sigma_{1}=\sigma_{0}$, $\sigma_{2}=\sigma_{\mathrm{p}}$).
First, we decrease the length of the left-hand-side negative region starting from '$\mathrm{nn}$', and increase the length of the right-hand-side neutral region ('$\mathrm{n}0$').
An intermediate limiting case is the fully neutral pore:  '$00$'.
Then, we increase the length of the right-hand-side positive region ('$0\mathrm{p}$') until we reach the limiting case '$\mathrm{pp}$'.
Note that subscripts $1/2$ are dropped at '$\mathrm{n}0$', '$00$', and '$0\mathrm{p}$'.

Each charge pattern can be characterized by a net pore charge.
We will use the dimensionless number, $Q$, ranging from $-1$ to $1$, that is obtained as 
\begin{equation}
 Q=x_{1}\dfrac{\sigma_{1}}{\sigma_{\mathrm{p}}} + (1-x_{1})\dfrac{\sigma_{2}}{\sigma_{\mathrm{p}}} .
\end{equation} 

We have performed MD simulations for the above charge patterns based on the all-atom model described in Section \ref{subsec:md}.
We produced whole current-voltage curves (data not shown).
According to our expectations, they are linear for symmetric pores ('$\mathrm{nn}$', '$\mathrm{pp}$', and '$00$'), while they exhibit rectification for asymmetric pores ('$\mathrm{np}$', '$\mathrm{n}0$', and '$0\mathrm{p}$').
We present results for $200$ and $-200$ mV (ON and OFF states), because these values are representative characterizing both device functions (rectification and selectivity).

\begin{figure}
\begin{center}
\includegraphics*[width=0.43\textwidth]{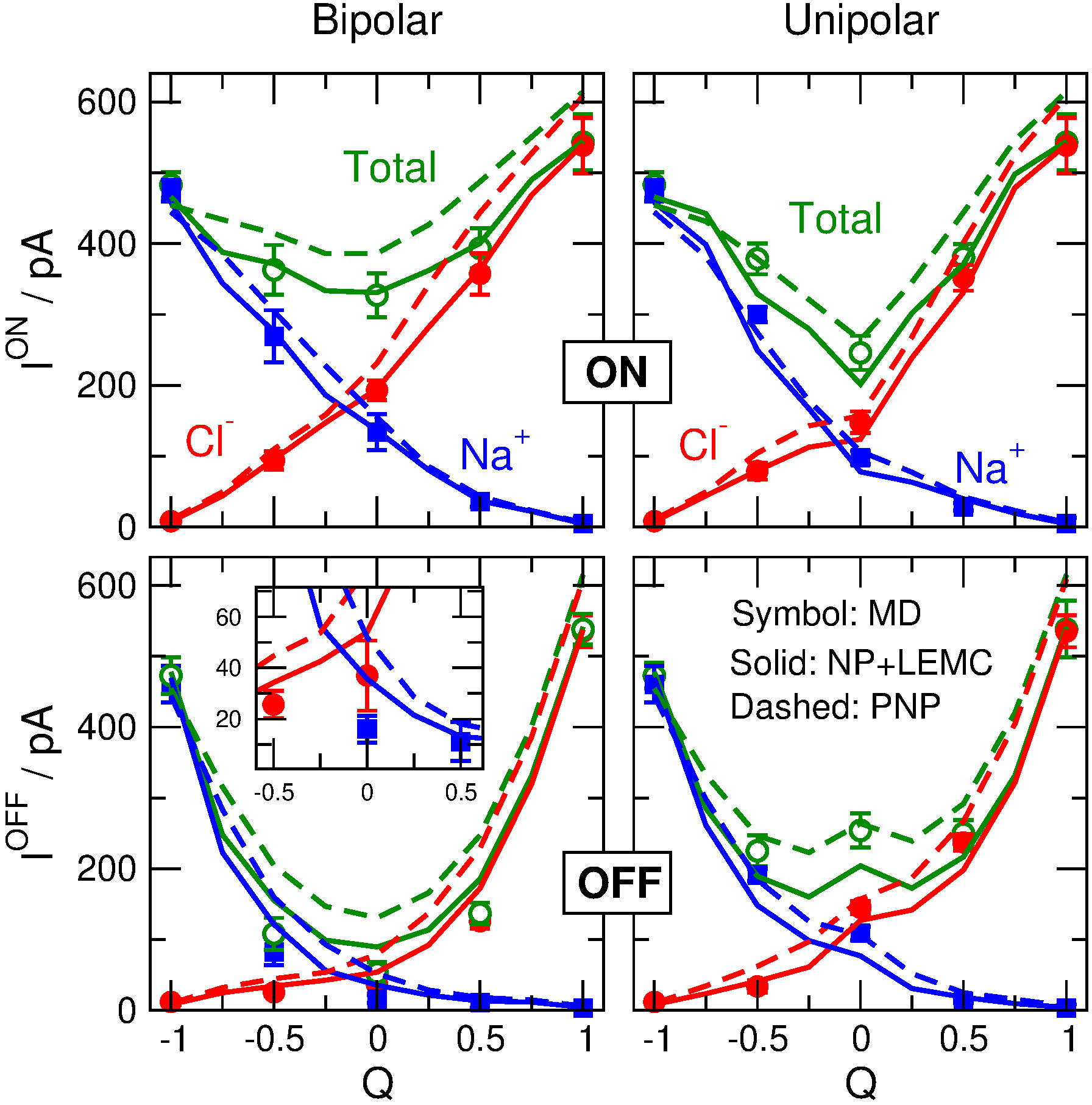}
\end{center}
\caption{
Total currents (green) and those carried by Na$^{+}$ (blue) and Cl$^{-}$ (red) as functions of the normalized total pore charge, $Q$, in the bipolar (left column) and unipolar (right panel) geometries.
Top row refers to the ON state ($200$ mV), while the bottom row refers to the OFF state ($-200$ mV) of the nanopore.
Results obtained with MD, NP+LEMC, and PNP are shown with symbols, solid lines, and dashed lines, respectively.
}
\label{fig3}
\end{figure}

The currents obtained from MD are plotted in Fig.\ \ref{fig3} as functions of $Q$, while concentration profiles are shown in Fig.\ \ref{fig4}.
MD data are shown with symbols with error bars, where the statistical error is larger than the size of the symbol.

\subsection{Fitting the diffusion coefficient in the pore}

Computation of NP+LEMC and PNP data requires fitting the diffusion coefficient profiles, $D_{i}(\mathbf{r})$, to MD data.
In our previous work \cite{hato_pccp_2017}, we published an extensive study for the '$\mathrm{np}$' bipolar nanopore comparing explicit-water (studied by MD) to implicit-water (studied by NP+LEMC) models.
There, the diffusion coefficient profile was a piecewise constant function along the $z$-axis: $D_{i}^{\mathrm{bath}}$ in the baths and $D_{i}^{\mathrm{pore}}$ inside the pore.
While we used experimental diffusion constant values for $D_{i}^{\mathrm{bath}}$, we fitted the $D_{i}^{\mathrm{pore}}$ value to MD current data in a single state point.
That state point was the ON state (200 mV) for $c=1$ M concentration and $\pm \sigma_{\mathrm{p}}$ surface charge. 
(Note that the value of $\sigma_{\mathrm{p}}$ was erroneously reported in Ref.\ \cite{hato_pccp_2017} as 1 $e$/nm$^{2}$; the correct value is 0.4835 $e$/nm$^{2}$ and is also used here.)

After obtaining the $D_{i}^{\mathrm{pore}}$ values for that state point, we fixed these values and used them for other voltages, concentrations, and surface charges in Ref.\ \cite{hato_pccp_2017}.
The main practical question of  that work from the point of view of multiscaling was to what degree these fitted $D_{i}^{\mathrm{pore}}$ values are transferable to other state points.
We found that transferability worked quite well, at least, qualitatively. 
The basic explanation is that changing conditions (voltage, concentration, surface charge) changes the axial concentration profiles, $c_{i}(z)$, similarly in NP+LEMC and MD.

The diffusion coefficient has a well-established physical meaning characterizing the ability of ions to diffuse in a given environment.
When we use it for connecting two fundamentally different models (explicit water vs.\ implicit water), however, it becomes an adjustable parameter whose primary job is to take into account the differences between the two models.
By model here we mean the interactions between the particles of the system that determine the potential energy, through which, they determine the concentration profiles.
Concentration profiles are the main raw results that reflect the differences between the two molecular models. 

The empirical NP equation (Eq.\ \ref{eq:np}) formally separates the availability of charge carriers ($c_{i}$), their mobility ($D_{i}$), and the driving force ($\nabla\mu_{i}$).
In our MD simulations, we have information only about the concentration profiles and the current 
\begin{equation}
I_{i}=q_{i}\int_{A}\mathbf{j}_{i}(\mathbf{r})\cdot \mathrm{d}\mathbf{a} ,
\label{eq:Ii}
\end{equation} 
where $\mathrm{d}\mathbf{a}=\mathrm{d}a\, \mathbf{k} $ with $\mathbf{k}$ being the unit vector in the $z$ direction and $A=R_{\mathrm{pore}}^{2}\pi $ is the cross section of the pore.
So, concentration profile and current are common points in the two methods.
If we want to design a procedure for fitting the diffusion coefficient, we need to examine the concentration profiles as obtained on the basis of the two models.
We need to judge the differences between the two models and how these differences change with conditions, so we can conclude at what conditions we need to fit the $D_{i}^{\mathrm{pore}}$ values and at what conditions they are transferable.

\begin{figure*}
\begin{center}
\includegraphics*[width=0.95\textwidth]{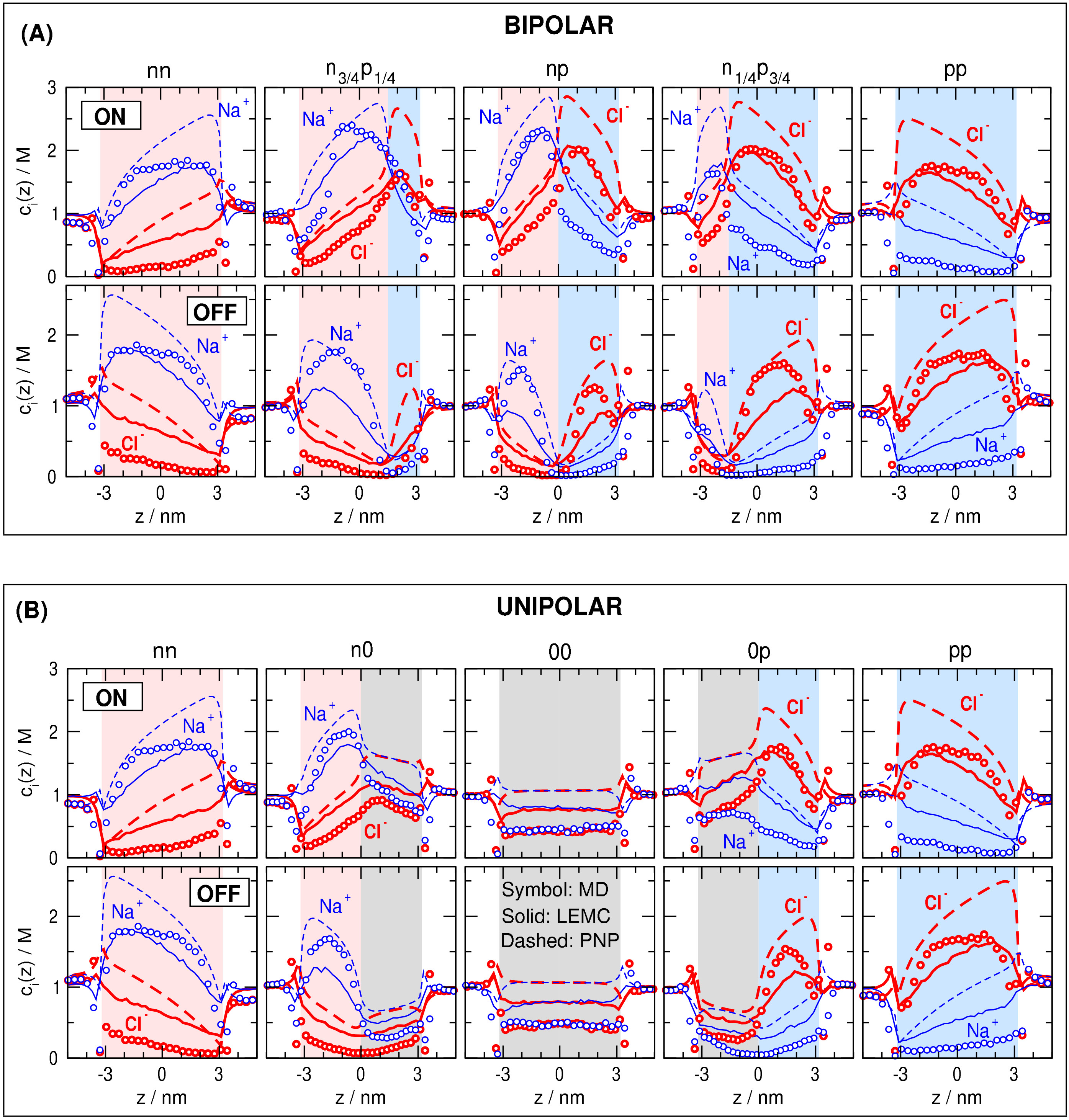}
\end{center}
\caption{
Concentration profiles of Na$^{+}$ (blue thin lines) and Cl$^{-}$ (red thick lines) ions for (A) the bipolar and (B) the unipolar cases. 
In both cases, top row and bottom row show the ON and OFF states, respectively.
Results obtained with MD, NP+LEMC, and PNP are shown with symbols, solid lines, and dashed lines, respectively.
Concentration profiles are computed  by normalizing for the same cross section ($A=R_{\mathrm{pore}}^{2}\pi$) in all the three methods although the effective cross sections through which ion centers traffic might be different in the three cases (see main text).
}
\label{fig4}
\end{figure*}

Let us look at Fig.\ \ref{fig4} that also shows the NP+LEMC and PNP concentration profiles as obtained using the $D_{i}^{\mathrm{pore}}$ values provided by the fitting procedure described in the following.
If we look at the Cl$^{-}$ profiles (red symbols and solid lines) in the top row, for example, we can see that NP+LEMC overestimates the MD profiles for the '$\mathrm{nn}$' geometry ($Q=-1$). 
As we proceed by adding more 'p' region to the right (increasing $Q$) the degree of this overestimation gradually decreases.
Therefore, there is a distinct correlation between the difference between the NP+LEMC and MD profiles (the difference that $D_{i}^{\mathrm{pore}}$ is supposed to balance) and pore charge ($Q$).
Looking at other cases (OFF states and unipolar pores), we observe a similar trend.

Therefore, we follow the practice of our previous work \cite{hato_pccp_2017} and fit the $D_{i}^{\mathrm{pore}}$ values in the ON state for the bipolar pore, but this time we perform the fitting for each $Q$ value.
Then, we fix these $D_{i}^{\mathrm{pore}}$ values and  examine whether they are transferable to the OFF states and to unipolar pores.
Also, we study whether the values obtained by the NP+LEMC method are transferable to another method, PNP.
Doing that we can also measure the accuracy of PNP.

\begin{figure}
\begin{center}
\includegraphics*[width=0.35\textwidth]{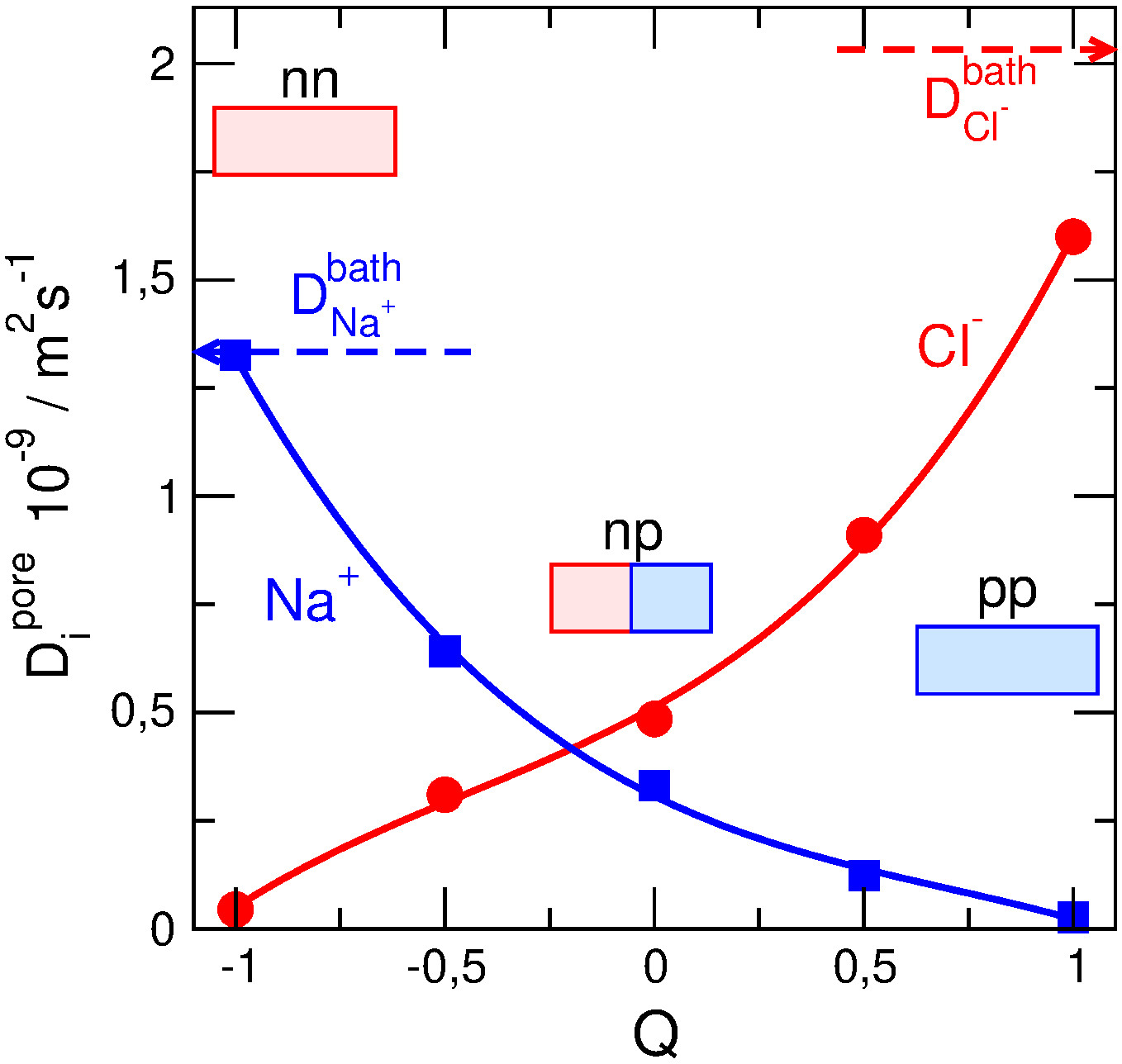}
\end{center}
\caption{Diffusion coefficients in the pore, $D_{i}^{\mathrm{pore}}$, obtained by fitting the currents computed from NP+LEMC to currents provided by MD for the same cases.
Fitting have been done for the bipolar cases in the ON state for different $Q$ values.
The obtained $D_{i}^{\mathrm{pore}}$ values were used in other cases (unipolar pores, OFF states, PNP) and transferability of these values investigated.
Bath values, $D_{i}^{\mathrm{bath}}$, are indicated by arrows.
The lines are fitted curves that provide the $D_{i}^{\mathrm{pore}}$ values for states (in terms of $Q$) for which MD simulation is not available.
}           
\label{fig5}
\end{figure} 

Figure \ref{fig5} shows the results for the fitted diffusion coefficients in the pore as functions of $Q$ for Na$^{+}$ and Cl$^{-}$. 
For $Q=-1$, Cl$^{-}$ is the coion. 
Its concentration profile is considerably overestimated by NP+LEMC, so its $D_{\mathrm{Cl}^{-}}^{\mathrm{pore}}$ value is small compared to the bath value in order to balance the overestimation.
For $Q=1$, the same is true for the Na$^{+}$ ion. 
In between a smooth behavior is observed as judged previously from the concentration profiles. 

Figure \ref{fig5} reveals that fitting works somewhat differently for Na$^{+}$ and Cl$^{-}$.
We need to decrease the diffusion coefficient of Cl$^{-}$ compared to the bath value more than we need to do that for Na$^{+}$.
We hypothesize that this is due to the different behavior of these ions with respect to their hydration shells.
In bulk, diffusivity of Cl$^{-}$ is larger, because its hydration shell, at least according to textbook explanations, is less stable, so its hydrodynamic radius is smaller.

Entering the pore, the stability of hydration shells is partially lost due to confinement and strong local electric fields, so this comparative advantage of Cl$^{-}$ with respect to Na$^{+}$ is lost.

\subsection{Transferability to PNP}

If we use these $D_{i}^{\mathrm{pore}}$ values in the NP+LEMC and PNP methods, we obtain the currents plotted by solid and dashed lines, respectively, in Fig.\ \ref{fig3}.
In the top-left panel (bipolar, ON state), the NP+LEMC curves (solid) agree with the MD points, because these points form the basis of the diffusion coefficient fit.
The PNP results (dashed) overestimate the NP+LEMC results.
This is a general conclusion valid for all cases and its explanation is that the PNP concentration profiles are generally larger than either the NP+LEMC or the MD data (see Fig.\ \ref{fig4}).

In the PNP theory, ions are point charges ``feeling'' the mean electric field. 
Because of the lack of ion size, the effective cross section through which ion centers can cross the pore is larger in the case of PNP ($R_{\mathrm{pore}}^{2}\pi$) than in the case of NP+LEMC ($(R_{\mathrm{pore}}-R_{i})^{2}\pi$).
Also, the pore can take more ions in PNP due to the lack of hard sphere exclusion.
This can be seen most clearly in the '$00$' case (Fig.\ \ref{fig4}).

Because the deviations between NP+LEMC and PNP are relatively small, we can conclude that the diffusion coefficient values, $D_{i}^{\mathrm{pore}}$, obtained for NP+LEMC are fairly transferable to PNP.
After stating this for PNP, we will consider only NP+LEMC in comparison of the implicit-water model to the explicit-water model (MD) in discussions from now on.

\subsection{Transferability to unipolar pores}

If we compare the bipolar case with the unipolar case in the ON state, we can observe that even MD provides quite similar currents in these two cases for a given $Q$.
This is especially strange for $Q=0$, because the '$\mathrm{np}$' and the '$00$' pores are strikingly different just considering their charge patterns.
This difference  is also reflected by the concentration profiles (Fig.\ \ref{fig4}).

To explain this, we need to return to the concentration profiles, to the quantity that is provided by both MD and NP+LEMC.
Looking at just the $c_{i}(z)$ profiles in Fig.\ \ref{fig4}, it is \emph{not} clear right away, why the '$\mathrm{np}$' and the '$00$' pores let so similar currents through.
For a better understanding, we need to consider the analysis that we used some time ago to study currents through calcium channels on the basis of equilibrium GCMC simulations\cite{gillespie_bj_2008,gillespie_bj_2008b,boda_jgp_2009,malasics_bba_2010}. 

That analysis approximated the slope conductance (reciprocal of the resistance) that is defined for ionic species $i$ as
\begin{equation}
 g_{i} = \left(\dfrac{\mathrm{d}I_{i}}{\mathrm{d}U} \right)_{U=0} ,
\end{equation} 
where the ionic current, $I_{i}$, is defined in Eq.\ \ref{eq:Ii}, and $U$ is the voltage.
Using the NP equation for $\mathbf{j}_{i}(\mathbf{r})$, assuming that the chemical potential is constant in the radial dimension ($\mu_{i}(z,r)\approx \mu_{i}(z)$) and that the bottleneck for the ion transport is the pore (the pore is the highest-resistance element), we can write that 
\begin{eqnarray}
 I_{i} & \approx & -\dfrac{q_{i}D_{i}^{\mathrm{pore}}}{kT}  \dfrac{\mathrm{d}\mu_{i}(z)}{\mathrm{d}z} \int\limits_{0}^{R_{\mathrm{pore}}} c_{i}(z,r) 2\pi r \mathrm{d} r \nonumber \\
 & = & -\dfrac{q_{i}D_{i}^{\mathrm{pore}}}{kT} \dfrac{\mathrm{d}\mu_{i}(z)}{\mathrm{d}z} A c_{i}(z)  .
\end{eqnarray} 
Solving for $\mathrm{d}\mu_{i}(z)/\mathrm{d}z$ on the right hand side and integrating along the pore from $-H/2$ to $H/2$, we obtain
\begin{eqnarray}
&-\dfrac{kT I_{i}}{q_{i}AD_{i}^{\mathrm{pore}} } \int\limits_{-H/2}^{H/2} \dfrac{\mathrm{d}z}{c_{i}(z)}  =  \int\limits_{-H/2}^{H/2}  \dfrac{\mathrm{d}\mu_{i}(z)}{\mathrm{d}z} \mathrm{d}z  \nonumber \\
&= \Delta \mu_{i} = q_{i}U ,
\end{eqnarray} 
where $\Delta  \mu_{i}=q_{i}U$ is the drop of the electrochemical potential across the pore using the fact that the electrolyte concentrations are the same at the two sides of the membrane.
From this equation, the resistance can be approximated as
\begin{equation}
 g_{i}^{-1} = \dfrac{U}{I_{i}} = -\frac{kT}{q_{i}^{2}AD_{i}^{\mathrm{pore}} } \int\limits_{-H/2}^{H/2} \dfrac{\mathrm{d}z}{c_{i}(z)} ,
 \label{eq:resistance}
\end{equation} 
if $U$ is small enough.
The resistance of the pore, therefore, can be associated with the integral under $c_{i}^{-1}(z)$.

Figure \ref{fig6} shows the reciprocal of the concentration profiles of Fig.\ \ref{fig4} for $Q=0$ (only the MD and NP+LEMC data).
The top row shows the results for the ON state.
For the unipolar pore, the $c_{i}^{-1}(z)$ profile is closely constant with the area under the curves being proportional to the resistance.
The curves for the bipolar pore, on the other hand, show variations, because an ion has large concentration (peak) in the region where it is a counterion, while it has small concentration (depletion zone) in the other region where it is a coion.

Different regions of the nanopore along the $z$-axis can be considered as resistors connected in series.
While the two (left and right) regions of the unipolar pore have the same intermediate resistances, the left and right regions of the bipolar pore  have different resistances.
In the integral of Eq.\ \ref{eq:resistance} they add up, so regions of low and high resistances tend to balance each other.
That is why the structurally different unipolar and bipolar pores have similar resistances (currents) as given by MD.

As far as the transferability of the diffusion coefficient, $D_{i}^{\mathrm{pore}}$, is concerned, Fig.\ \ref{fig6} shows numbers indicating the ratios of the integrals of the NP+LEMC and MD curves.
This ratio is similar for the bipolar and unipolar pores (0.57 vs.\ 0.56 for Na$^{+}$ and 0.75 vs.\ 0.57 for Cl$^{-}$).
This means that the integral can be corrected (note the presence of $D_{i}^{\mathrm{pore}}$ in Eq.\ \ref{eq:resistance}) with the same scaling factor in the bipolar and the unipolar cases.

\subsection{Transferability to the OFF state}

If we compare the currents between the ON and OFF states for the bipolar case (left column of Fig.\ \ref{fig3}), we can judge that currents in the OFF state as computed by NP+LEMC with the transferred $D_{i}^{\mathrm{pore}}$ values agree qualitatively well with MD data. 
This agrees with our finding in our previous work \cite{hato_pccp_2017}: ions respond similarly to the changing sign of voltage in the explicit- and implicit-water models.

If we look at the results more closely, however, we can observe a weaker agreement between the MD and NP+LEMC currents than in the bipolar vs.\ unipolar comparison.
This is well visible in the inset of the bottom-left panel of Fig.\ \ref{fig3} that shows that the NP+LEMC data are 1.5--2 times larger than the MD data.

\begin{figure}
\begin{center}
\includegraphics*[width=0.43\textwidth]{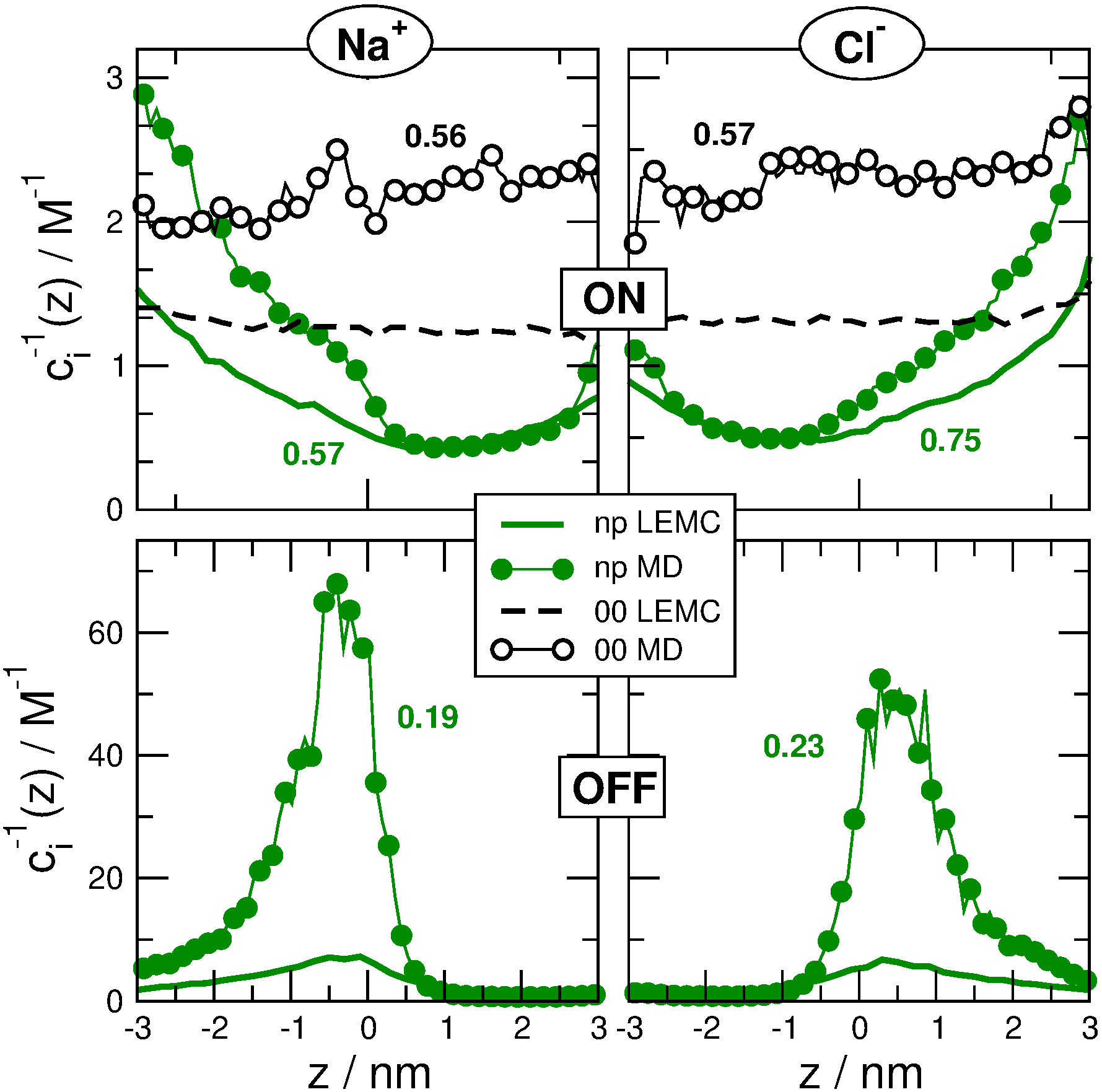}
\end{center}
\caption{Reciprocal concentration profiles of Na$^{+}$ (left) and Cl$^{-}$ (right) for $Q=0$ as obtained from MD (symbols) and NP+LEMC (lines). Top and bottom rows refer to the ON and OFF states, respectively.
Green filled symbols and solid lines refer to the bipolar pore ('$\mathrm{np}$'), while black open symbols ans dashed lines refer to the unipolar pore ('$00$').
The curves for the '$00$' geometry are not shown in the OFF state, because they are not visible on the scale of the ordinate; they are the same as in the ON state. 
The numbers indicate the ratios of the integrals of the NP+LEMC and MD curves.
}
\label{fig6}
\end{figure}

The explanation can be deduced from Fig.\ \ref{fig6} by comparing the ON and OFF states.
The depletion zones of the MD profiles are much deeper in the OFF state than the depletion zones of the NP+LEMC profiles.
This means that the $c_{i}^{-1}(z)$ curves are much larger.
The integrals are also very different in the MD and NP+LEMC cases for the OFF state; the ratios of the integrals are 0.19 and 0.23 for Na$^{+}$ and Cl$^{-}$, respectively.
The fact that the transferability of the $D_{i}^{\mathrm{pore}}$ values is imperfect is shown by that these values are 0.57 and 0.75 for the ON state. 

The behavior of the axial concentration profiles, therefore, can account for the difference between the ON and OFF states only partially.
The diffusion coefficient as a scaling factor is also needed to take into account the difference between the explicit- and implicit-water models.
A $D_{i}^{\mathrm{pore}}$ value 1.5--2 times smaller would be necessary in the OFF state to reach agreement between MD and NP+LEMC.
Still, transferability of the $D_{i}^{\mathrm{pore}}$ values can be considered quite good if we take the profoundly different natures of the implicit- and explicit-water models.

\begin{figure}[t!]
\begin{center}
\includegraphics*[width=0.46\textwidth]{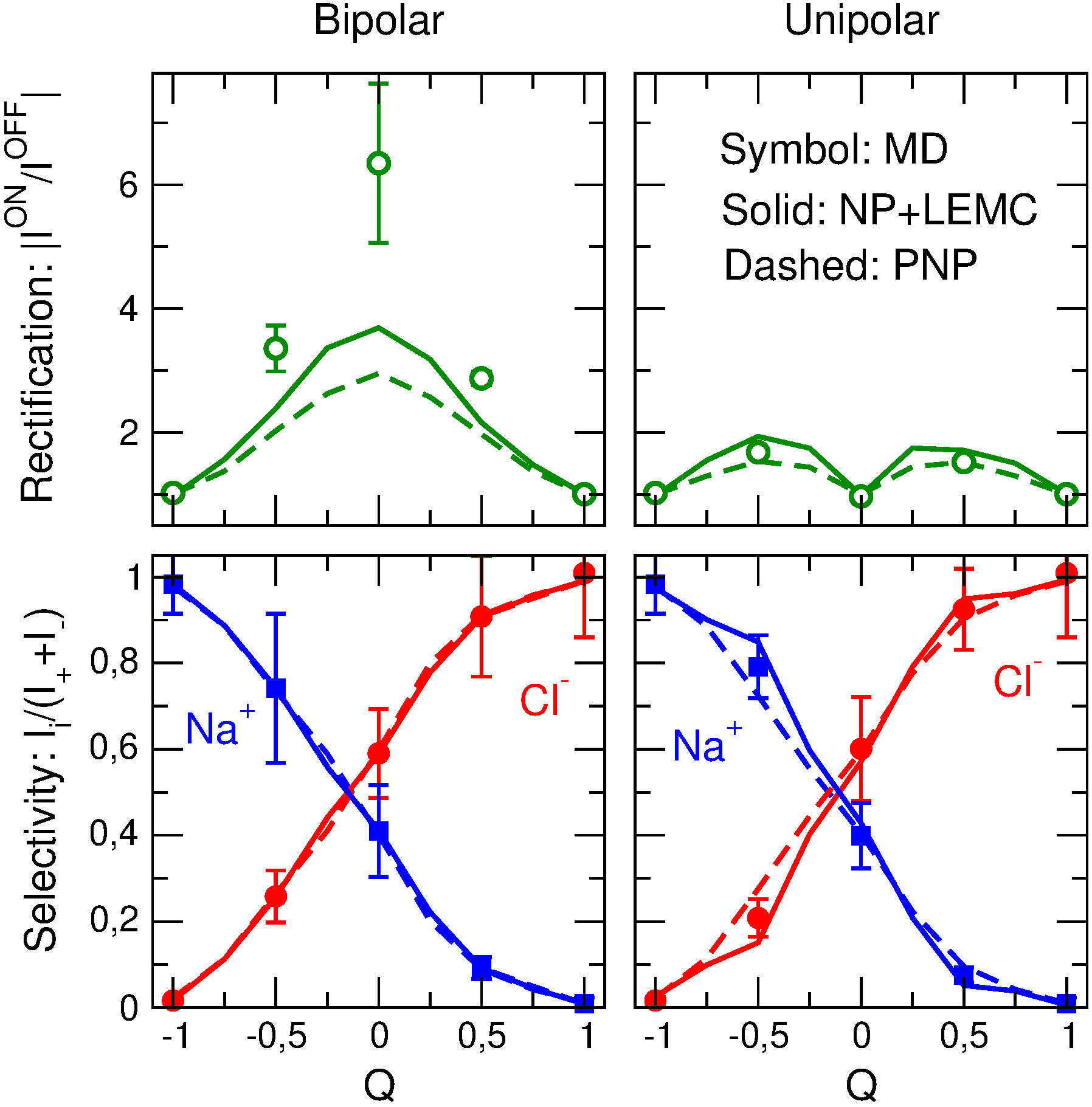}
\end{center}
\caption{Rectification (top) and ON-state selectivity (bottom) of the bipolar (left) and unipolar (right) pores as functions of $Q$.
Rectification defined as the absolute value of the ratio of the ON and OFF currents is defined in terms of total currents.
Selectivity defined as the fraction of current carried by ionic species $i$ is plotted for both ions.
Results obtained with MD, NP+LEMC, and PNP are shown with symbols, solid lines, and dashed lines, respectively.
}
\label{fig7}
\end{figure} 

\subsection{Rectification and selectivity}

The deviation between the MD and NP+LEMC OFF-currents is also apparent from Fig.\ \ref{fig7}, where the rectifications and the selectivities are plotted. 
Rectification is defined as the absolute value of the ratio of the currents in the ON and and OFF states: $|I^{\mathrm{ON}}/I^{\mathrm{OFF}}|$.
Selectivity is defined as the fraction of the total current carried by ionic species $i$: $I_{i}/I$, where $I=I_{+}+I_{-}$ is the total current.
The pore is selective for a given ion, if this value is close to $1$, while it is non-selective if the value is about $0.5$ (for a 1:1 electrolyte).

The figure shows that the nanopore can be turned from being selective to being rectifying by changing charge pattern characterized by the pore charge, $Q$.
The uniformly charged pores ($Q=-1$ or $Q=1$) are selective for the counterion, but they do not rectify.
As the pore becomes asymmetric, it becomes rectifying at the price of becoming less selective.

The bipolar pores rectify better, because they are more asymmetric with stronger polarity.
The maximum of rectification is around $Q=0$ that is the bipolar nanopore studied previously\cite{hato_cmp_2016,hato_pccp_2017,matejczyk_jcp_2017}.
The phenomenon of rectification was analyzed in these studies in detail.
Briefly, rectification occurs because the depletion zones of the coions are deeper in the OFF state due to the additional effect of the applied field.

The unipolar pores rectify more weakly with maxima around $Q=-0.5$ and $Q=0.5$.
The special case of $Q=0$ ('$00$')neither rectifies nor selective.
Their selectivity behavior is similar to that of bipolar pores, because selectivity primarily depends on $Q$.

\section{Summary}

The main message of Fig.\ \ref{fig7} is that the reduced model, studied either with NP+LEMC or PNP, is able to reproduce device functions for a wide variety of charge pattern; that is, reduced models are useful in understanding the mesoscale physics of the device (i.e., how the current is produced).  
Moreover, the effective diffusion coefficients are relatively transferable to different voltages and charge patterns.
In general, reduced models contain adjustable parameters with physical significance (here, diffusion coefficient in the pore) that can scale the system's device-level behavior and account for differences between the reduced model and the more realistic case (experiment or more detailed model). 
These parameters can be fairly transferable if the reduced model captures the basic device physics (here, axial concentration profiles).

MD simulation is useful if we are interested in molecular details beyond the implicit-solvent model.
There is, for example, a water layer between ions and pore wall as revealed by radial concentration profiles and contour plots (see Fig.\ 4 of Ref.\ \cite{hato_pccp_2017}).
Electrical potential distributions can also be studied to reveal how water molecules screen the charges of ions as opposed to the simple division by $\epsilon$ in the implicit-water case (see Figs.\ 5 and 6 of Ref.\ \cite{hato_pccp_2017} and Fig.\ 10 of Ref.\ \cite{hato_cmp_2016}).
If we are after ``just'' the device behavior, however, reduced models can be extremely handy.

A multiscaling approach is useful, because if we work with all the modeling levels in mind, we never lose sight of the drawbacks of the model that we are currently working with.
If we work with a reduced model, for example, we need to keep in mind that there are approximations and omissions in action.
We need to be aware the nature of these approximations and omissions, how do they influence the results, whether they affect device functions, and what are the contents of the adjustable parameters that are always present in reduced models.
To obtain answers to such questions, we need to relate our results from the reduced model to something more fundamental, MD results, for example.

If we work with all-atom models, we need to keep in mind the limitations of force fields, system sizes, and insufficient sampling.
We need to be aware that too many details can conceal relevant emergent patterns that can be revealed by reduced models because reduced models can concentrate on specific physics.
Specifically, after the reasonableness of the reduced model is established, it can be used to understand the device physics, which is generally easier than using only MD simulations.
There is a long way ahead of us until reliable force fields are developed even if computer power keeps increasing.
Until then, reduced models will continue to play a vital role.

PNP has also proved to be quite accurate, but this is due to the fact that we are dealing with a problem, where mean-field electrostatics has a primary effect on the behavior of concentrations, and, thus, device behavior.
Interaction with surface charge and applied field is captured reasonably well by PNP.

There are cases, however, when mean-field electrostatics is not sufficient any more as in the case of crowded ion channels with a strong competition between Na$^{+}$ and Ca$^{2+}$, for example, where both ion size and electrostatic correlations are of crucial importance \cite{2000_nonner_bj_1976,gillespie_bj_2008,gillespie_bj_2008b,boda_jgp_2009,malasics_bba_2010}. 
Divalent and trivalent ions, as well as large monovalent ions, are also problematic.
Interesting phenomena applicable in nanofluidic devices can occur in such cases that are not reproduced by PNP.
One notable example of these phenomena is charge inversion \cite{he-jacs-131-5194-2009,loessbergzahl_ac_2016,chou_nl_2018} and ion layering \cite{gillespie_nl_2012}.

\section*{Acknowledgements}
\label{sec:ack}

We gratefully acknowledge  the financial support of the National Research, Development and Innovation Office -- NKFIH K124353. 
Present article was published in the frame of the project GINOP-2.3.2-15-2016-00053 (``Development of engine fuels with high hydrogen content in their molecular structures (contribution to sustainable mobility)'').
BM acknowledges financial support from the Austrian Academy of Sciences \"OAW via the New Frontiers Grant NST-001.

%

\end{document}